\documentclass[aps,prl,twocolumn,showpacs,preprintnumbers,amsmath,amssymb]{revtex4-1}

\usepackage{graphicx}
\usepackage{bbold}

\begin{document}
\title{Inhomogeneities and impurities in a dense one-dimensional Rydberg lattice gas}
\author{S. Ji, V. Sanghai, C. Ates, and I. Lesanovsky}
\affiliation{Midlands Ultracold Atom Research Centre (MUARC), School
of Physics and Astronomy, The University of Nottingham, Nottingham,
NG7 2RD, United Kingdom}

\begin{abstract}
We consider a dense one-dimensional laser-driven Rydberg lattice gas with perfect nearest-neighbor blockade. The ground state of this system can be found analytically in certain parameter regimes even when the applied fields are inhomogeneous in space. We will use this unique feature to investigate the effect of an impurity - introduced by the local variation of the laser parameters - on the correlations of the many-body ground state. Moreover, we explore the role of a staggered laser field which alternates from site to site thereby breaking the sublattice symmetry. We demonstrate that this technique, which can be applied experimentally, reveals insights into the role of  long-range interactions on the critical properties of a Rydberg gas. Our work
highlight novel possibilities for the exploration of many-body physics in Rydberg lattice gases based on locally tuneable laser fields.
\end{abstract}

\maketitle
Lattice gases of Rydberg atoms are currently in the focus of intense research as a platform to investigate strongly-correlated quantum systems. A host of theoretical works has studied and analyzed many-body phenomena such as the formation of crystalline structures as well as their melting under the influence of quantum fluctuations or dissipation \cite{atpo+:07,wehe+:08,bism+:11,pode+:10,webu:10,scle+:10,le:11,*le:12,leka:12,sepu+:11,zema+:12,leha+:12,atga+:12,tata+:12,gahe:+12,gale:12,homu+:13}. These investigations have revealed detailed insights into the equilibrium properties and non-equilibrium phenomena of strongly interacting many-body quantum systems.

On the experimental side there has recently been much progress. While initial experiments had revealed first evidence for the strong and coherent interactions present among Rydberg atoms \cite{sire+:04,tofa+:04,gami+:09,urjo+:09,viba+:11,scsa+:12} a very recent experimental breakthrough has enabled the unambiguous identification of interaction effects manifesting themselves in strong spatial correlations among Rydberg atoms excited from a large lattice \cite{scch+:12}. This new class of experiments permits not only the direct visualization of Rydberg correlations but also the manipulation of external fields on the scale of individual lattice sites. This opens up new opportunities for the study of many-body phenomena and the exploration of phase transitions in Rydberg lattice gases.

In this work we perform a theoretical investigation of the ground state of a one-dimensional Rydberg lattice gas in the presence of inhomogeneities which are induced by spatial variations of external laser fields. On the one hand these spatial variations are intrinsic to any experiment and therefore an understanding of their effect is of technical relevance. On the other hand deliberately imposed inhomogeneities can be used to experimentally tackle important problems in condensed matter physics, such as the investigation of impurity phenomena and the detailed study of statics and dynamics of many-body quantum systems exposed to symmetry breaking fields. We show that despite the presence of such spatial inhomogeneities an approximate analytical description of the Rydberg lattice gas in terms of a frustration-free Hamiltonian can be obtained. We will use this feature to perform a detailed investigations of the effect of an impurity and alternating laser fields which break the sublattice symmetry of the Rydberg gas Hamiltonian. This work complements current studies on Rydberg gases which are almost exclusively conducted on homogeneous systems \cite{webu:10,scle+:10,sepu+:11,le:11,*le:12,leha+:12,zema+:12,homu+:13}. It illustrates the usefulness of the frustration-free description and sheds light on the critical properties of the ground state of a dense Rydberg lattice gas.

\begin{figure}
\includegraphics[width=1\columnwidth]{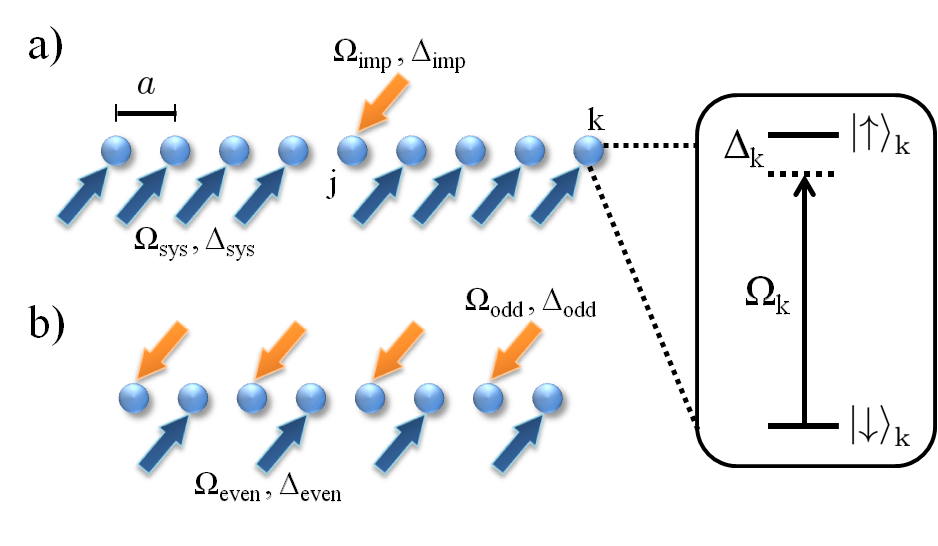}
\caption{Schematics of the one-dimensional lattice with spacing, $a$. Each site contains a single atom whose ground state $\left|\downarrow\right>_k$ is coupled to a Rydberg $nS$-state $\left|\uparrow\right>_k$ via a laser parameterized by a site-dependent Rabi-frequency $\Omega_k$ and detuning $\Delta_k$. (a) Rydberg lattice gas with a single impurity present on the $j$-th site. The impurity atom is irradiated by a laser (orange) of different parameters ($\Omega_{\text{imp}}, \Delta_{\text{imp}}$) compare to the rest of the system (blue, and parameterized by $\Omega_{\text{sys}}, \Delta_{\text{sys}}$). (b) Alternating lasers are introduced to investigate the effect of sublattice symmetry breaking fields. On odd (even) lattice sites, a laser parameterized by $\Omega_{\text{odd}}$ ($\Omega_{\text{even}}$) and  $\Delta_{\text{odd}}$ ($\Delta_{\text{even}}$) is used to excite atoms to Rydberg states.}
\label{fig:system}
\end{figure}

We consider a system in which atoms are held in a deep one-dimensional optical lattice with inter-site spacing $a$ (see Fig. \ref{fig:system}) and with a single atom per site, i.e., the external atomic degrees of freedom form a Mott-insulator. To model the internal dynamics we employ the well-established two-level pseudo-spin description: The electronic ground state of the atom located at the $k$-th lattice site is denoted by $\left|\downarrow\right>_k$. This state is coupled to a Rydberg n$S$-state, denoted by $\left|\uparrow\right>_k$, through a laser of Rabi frequency $\Omega_k$ and detuning $\Delta_k$, both of which can be site-dependent. Atoms in Rydberg n$S$-states interact via the van-der-Waals interactions $V(r)=C_6/r^6$ where $C_6$ is the dispersion coefficient and $r$ the interatomic distance. With these preliminaries we can now formulate the Hamiltonian of this many-body system on a lattice with $L$ sites:
\begin{equation}
H_{\text{Ryd}} =\sum^L_{k}  \Omega_k \sigma_{k}^x +  \sum^L_{k} \Delta_k n_{k} + \sum^L_{m>k} V_{|k-m|} n_{k}n_{m}.
\label{eqn:Ryd_Hamiltonian}
\end{equation}
Here $n_{k}=(1+\sigma_k^z)/2=\left|\uparrow\right>_k\!\!\left<\uparrow\right|$ is the projector on the Rydberg state and $\sigma_k^x$ and $\sigma_k^z$ are Pauli spin matrices. The interaction strength is parameterized by the constants $V_k=C_6/(ak)^6$.

In the following we will show that this Hamiltonian permits an approximate analytic solution over a certain range of parameters. To this end we generalize a construction put forward in Refs. \cite{le:11,*le:12} and consider a regime in which the laser Rabi-frequency is much weaker than the interaction among nearest neighbors, i.e. $V_1\gg\Omega_k$. Here the simultaneous excitation of neighboring atoms to Rydberg states is energetically forbidden. This suppression is made manifest in the Hamiltonian $H_\text{Ryd}$ by projecting it onto the subspace in which excited neighboring spins are absent, $H=\mathcal{P}\,H_\text{Ryd}\,\mathcal{P}$, using the projection operator $\mathcal{P}=\prod_k (1-n_k n_{k+1})$. Only the first term of $H_\text{Ryd}$ is affected by this projection. It transforms according to $\mathcal{P}\,\sigma^x_k\,\mathcal{P}=P_{k-1}\,\sigma^x_k\,P_{k+1}$,
where $P_k=1-n_k$ is the projector on the electronic ground state of the $k$-th atom.

We proceed by neglecting, for the moment, interactions beyond next-nearest neighbors, i.e. we set $V_k\rightarrow 0$ for $k>2$. Within these approximations the projected Hamiltonian of the Rydberg gas assumes the form $H=- \sum_{k}^L \Omega_k\xi_k+H_{\text{RK}}$, with
\begin{align}
H_{\text{RK}} =\sum_k h_k =
  \sum_{k}^L\Omega_k P_{k-1} \left( \sigma_k^x
 +\xi_k P_k + \xi_k^{-1} n_k\right)P_{k+1},
 \label{eqn:R-K Hamiltonian}
\end{align}
provided that the laser parameters and the interaction strength obey the relationship:
\begin{eqnarray}
\Omega_{k}= V_2\,\xi_k^{-1}&,\quad &\Delta_{k}= \Omega_k \left(\xi_k^{-1}-3\xi_k\right). \label{eqn:RK-manifold}
\end{eqnarray}
This can be checked by direct inspection utilizing the fact that one is working in the subspace $\mathcal{P}$ where $n_k n_{k+1}=0$. Eqs. (\ref{eqn:RK-manifold}) describe a region in parameter space where the projected Rydberg Hamiltonian is a so-called frustration-free or Rokhsar-Kivelson Hamiltonian \cite{roki:88,clcl+:05}. Here the ground state $\left|\{\xi_n\}\right>$, which depends on the set of parameters $\{\xi_n\}$, is annihilated by all local Hamiltonians $h_k$ such that $h_k\left|\{\xi_n\}\right>=0$. It can be constructed analytically and its explicit form reads
\begin{equation}
\left|\{\xi_n\}\right> = \frac{1}{\sqrt{Z_{\{\xi_n\}}}}\exp\left(-\sum_k^L \xi_k P_{k-1} \sigma_k^+ P_{k+1}\right)\left|0\right>,
\label{eqn:Ground state}
\end{equation}
where $Z_{\{\xi_n\}}$ is a normalization constant, $\left|0\right>=\left|\downarrow\downarrow \downarrow\cdots \downarrow \right>$ is the spin vacuum and $\sigma_k^+=(\sigma^x_k+i\sigma_k^y)/2=\left|\uparrow\right>_k\!\!\left<\downarrow\right|$. The relationship between the parameters $\xi_k$ and the laser parameters, i.e. $\Omega_k$, $\Delta_k$ , is obtained by eliminating $V_2$ from Eqs. (\ref{eqn:RK-manifold}):
\begin{equation}
\xi_k = \frac{1}{6}\left[-\frac{\Delta_k}{\Omega_k}+\sqrt{\left(\frac{\Delta_k}{\Omega_k}\right)^2+12}\right]. \label{eqn:xi}
\end{equation}
The square of $\xi_k$ can be interpreted as local fugacity since the probability for a spin on the $k$-th site to be in the state $\left|\uparrow\right>_k$ is proportional to $\xi_k^2$. This can be directly seen by expanding Eq. (\ref{eqn:Ground state}) in the spin product basis. In the following we will therefore refer to $\xi_k$ as fugacity parameter.

Let us now make use of the analytical knowledge of the ground state and investigate a Rydberg lattice gas in the presence of an impurity. In this scenario, which is depicted in Fig. \ref{fig:system}(a), all atoms but the $j$-th are irradiated by the same laser field characterized by $\Omega_\text{sys}$ and $\Delta_\text{sys}$ and the corresponding fugacity parameter $\xi_\text{sys}$, satisfying the relation (\ref{eqn:RK-manifold}). The impurity is introduced on site $j$ by applying here a laser field with in general different parameters given by $\Delta_\text{imp}$ and $\Omega_\text{imp}$. The corresponding fugacity parameter is then $\xi_\text{imp}$, which in an experimental setting can be achieved by single site addressing as experimentally demonstrated in Ref. \cite{scch+:12}.

\begin{figure}
\includegraphics[width=0.85\columnwidth]{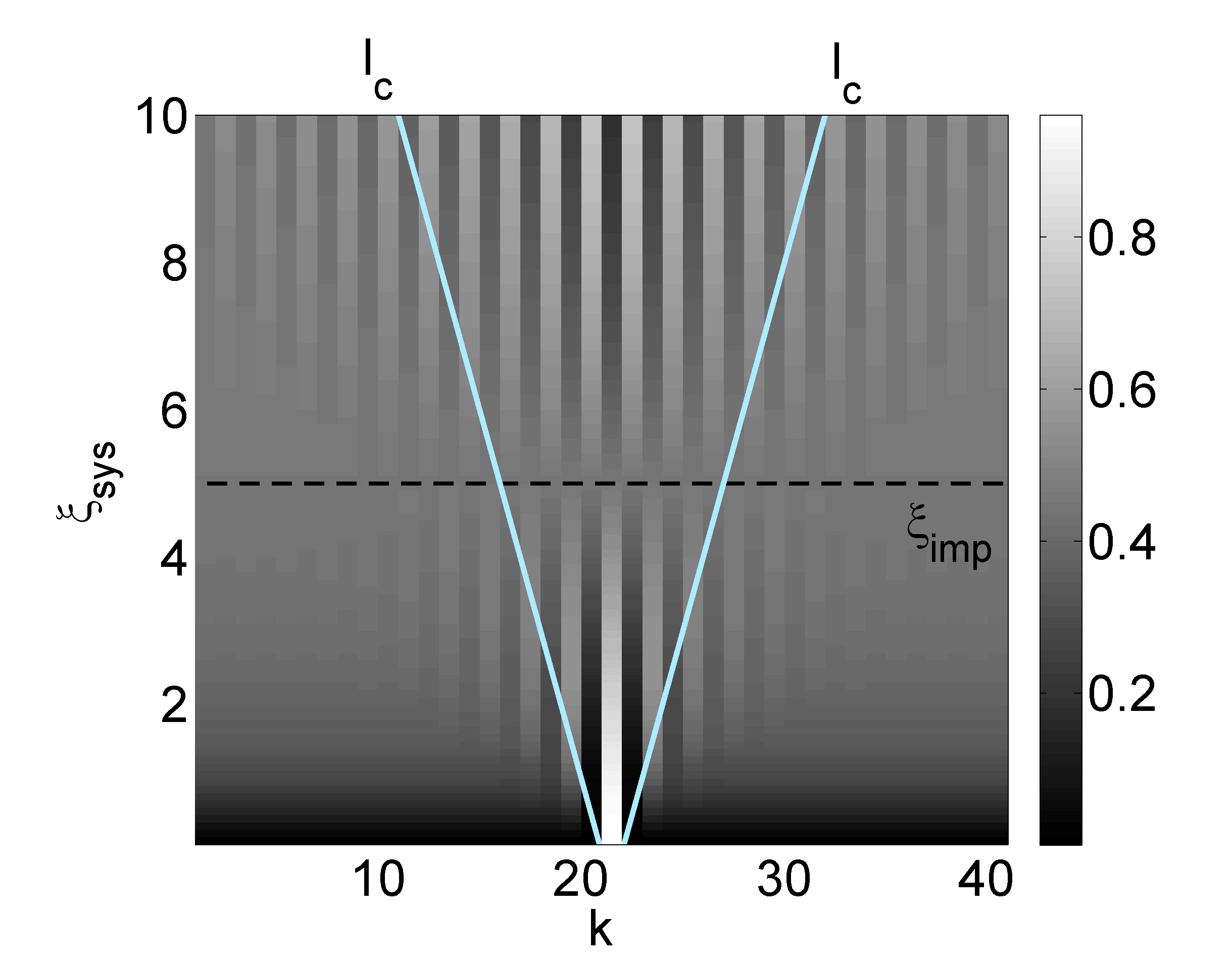}
\caption{Rydberg density for a lattice with $41$ sites and an impurity placed at site $j=21$. The fugacity parameter of the impurity is given by $\xi_\text{imp}=5$ (black dash) and the fugacity parameter of the remaining spins $\xi_\mathrm{sys}$ is varied from $0$ to $10$. With increasing $\xi_\mathrm{sys}$ the system's correlation length $l_c$ (plotted at the cyan diagonal lines) increases. The many-body state of the system atoms becomes more strongly correlated and the presence of the impurity affects the state of more and more distant atoms. Note that for $\xi_\mathrm{sys}=\xi_\mathrm{imp}$ the density is homogeneous in space due to the translational symmetry of the system.}
\label{fig:impurity}
\end{figure}
In Fig. \ref{fig:impurity} we show the spatial density distribution calculated for a lattice with periodic boundary conditions and length $L=41$. The impurity is introduced at site $j=21$ and its fugacity parameter is set to $\xi_\text{imp}=5$. The fugacity parameter of the system atoms $\xi_\text{sys}$ is varied on the vertical axis. For small $\xi_\text{sys}$ a peak of high density is visible at the site of the impurity which is expected since $\xi_\text{imp}>\xi_\text{sys}$ and thus, according to Eq. (\ref{eqn:Ground state}), the excitation probability at site $j$, is enhanced. When $\xi_\text{sys}$ is increased the overall density increases and density oscillations in the vicinity of the impurity indicate the onset of stronger and stronger correlations in the ground state. Indeed one can show that the correlation length of the system spins is given by $l_c=\xi_\mathrm{sys}$ \cite{le:11,*le:12}. This length is plotted as the cyan diagonal lines in the figure showing that the impurity indeed determines the state of the system atoms located within a distance $l_c$. In the vicinity of $\xi_\text{imp}=\xi_\text{sys}$ the density modulations vanish as here the system becomes translationally invariant.

When the fugacity parameter $\xi_\text{sys}$ is increased towards infinity - a limit which is achieved when $\Omega_\text{sys}$ approaches zero - the correlation length diverges. Here the ground state (\ref{eqn:Ground state}) of the homogeneous system, i.e. if the impurity was absent, becomes a superposition of the two anti-ferromagnetic states $\left|\uparrow\downarrow\uparrow\downarrow\uparrow...\right>$ and $\left|\downarrow\uparrow\downarrow\uparrow\downarrow...\right>$. Moreover, one can show that the excitation gap of the Hamiltonian (\ref{eqn:R-K Hamiltonian}) closes and in fact the impurity breaks the sublattice symmetry of the Hamiltonian (\ref{eqn:R-K Hamiltonian}) by forcing the ground state into one of the anti-ferromagnetic states with a down-spin at the impurity site.

To investigate the symmetry breaking region in parameter space in detail we introduce a staggered laser field. In an experimental setting this would be done by choosing laser parameters such that they give rise to a fugacity parameter $\xi_\text{odd}$ $(\xi_\text{even})$ at all odd (even) lattice sites, c.f. Fig. \ref{fig:system}(b). For the upcoming analysis it is convenient to define the difference $\xi_d=\xi_\text{odd}-\xi_\text{even}$ and the sum $\xi_s=\xi_\text{odd}+\xi_\text{even}$ of the two sublattice fugacity parameters.
\begin{figure}
\includegraphics[width=1\columnwidth]{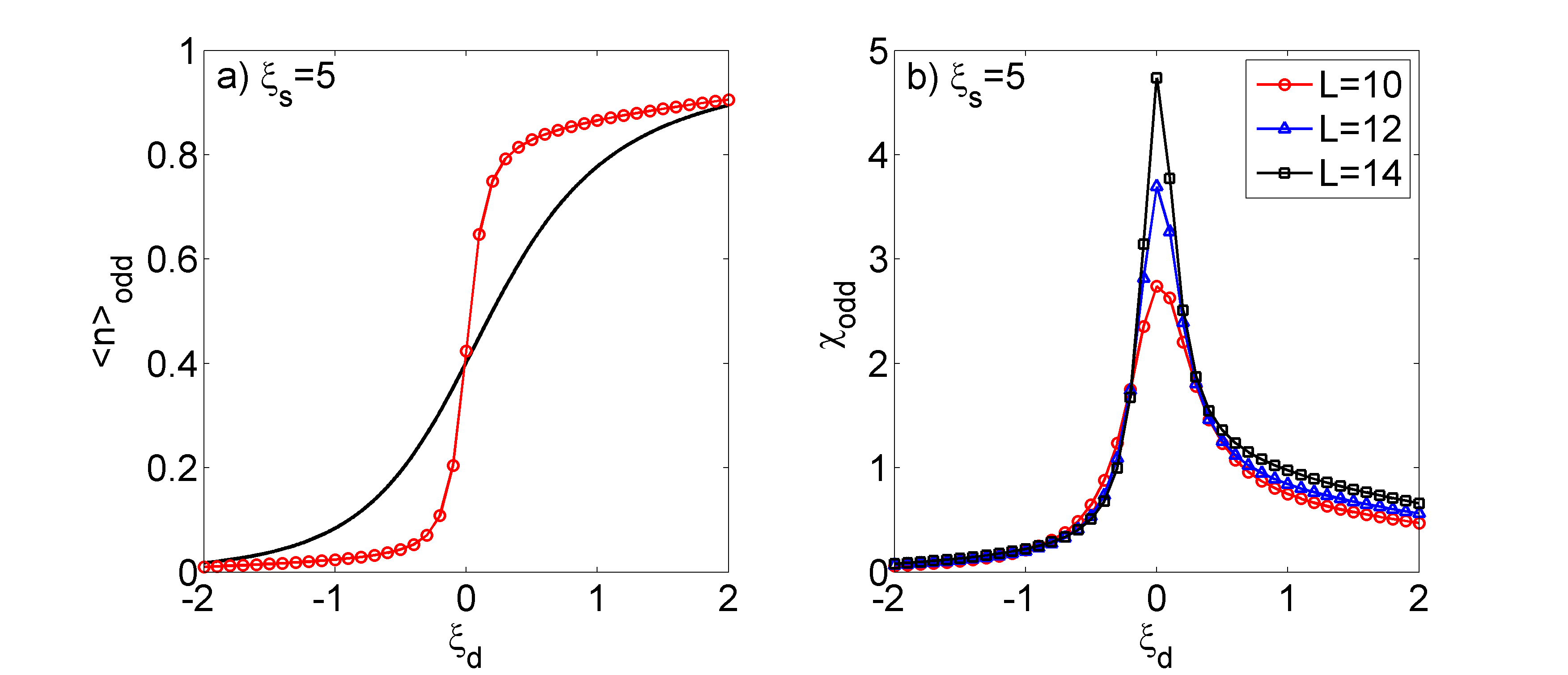}
\caption{(a) Mean density of the odd sublattice as a function of $\xi_d$. Here, with $\xi_s=5$, we have plotted the analytical result given in Eq. (\ref{eq:sub_lattice_density}) in black and the numerical result obtained from diagonalizing Hamiltonian (\ref{eqn:Ryd_Hamiltonian}) in red (with circles) with $L=14$. The latter shows a significantly steeper switching of the sublattice populations at $\xi_d = 0$. (b) Susceptibility $\chi_{\text{odd}}(\xi_s=5, \xi_d)$ for different lattice sizes: $L=10$ (red circles), $L=12$ (blue triangles), and $L=14$ (black squares). The data suggests a divergence of the susceptibility at $\xi_d=0$ in the limit of large lattice sizes $L$.}
\label{fig:transition}
\end{figure}
To investigate the effect of the symmetry breaking field let us now specifically study the expectation value of the Rydberg density on the odd sublattice $n_\mathrm{odd}=\sum_{k=\text{odd}} n_k$. Using the ground state (\ref{eqn:Ground state}) one finds,
\begin{eqnarray}
  \langle n_{\text{odd}} \rangle = \frac{1}{2}\left[1+\frac{\xi_d-\xi^{-1}_s}{\sqrt{(1+\xi^{-2}_s)(1+\xi^2_d)}}\right] \label{eq:sub_lattice_density}.
\end{eqnarray}
For small $\xi^{-1}_s$ this function predicts a transition between two states in which Rydberg atoms predominantly occupy the odd/even sublattice which takes place when the difference between the sublattice fugacity parameters vanishes $\xi_d=0$ [see Fig. \ref{fig:transition}(a)]. This is expected since for $\xi^{-1}_s=0$ and $\xi_d=0$ both $\left|\uparrow\downarrow\uparrow\downarrow\uparrow...\right>$ and $\left|\downarrow\uparrow\downarrow\uparrow\downarrow...\right>$ are ground states and any non-zero value of $\xi_d$ will favor one over the other.
Note, that according to Eq. (\ref{eq:sub_lattice_density}) this switching between the sublattices is not sharp as the susceptibility, i.e. the slope $\chi_{\text{odd}}(\xi_s,\xi_d)=\partial \langle n_{\text{odd}} \rangle/\partial \xi_d$, saturates at a value $1/2$ at the "transition point" $\{\xi_d=0, \xi_s^{-1} =0\}$ [see Eq.(\ref{eq:sub_lattice_density})]. Since a tiny perturbation to the system would lead to a symmetry breaking, e.g. an impurity, one would naturally expect the transition occurring at $\xi_d=0$ to be sharp rather than a crossover. Therefore, although the frustration-free Hamiltonian (\ref{eqn:R-K Hamiltonian}) excellently describes the Rydberg gas along the curve parameterized by Eqs. (\ref{eqn:RK-manifold}) as shown in Ref. \cite{le:11}, it is very questionable whether this Hamiltonian and Eq. (\ref{eq:sub_lattice_density}) faithfully describes the actual sublattice occupation of the ground state of the Rydberg gas Hamiltonian (\ref{eqn:Ryd_Hamiltonian}) at the "transition point". The suspicion is confirmed by numerically calculating $\langle n_{\text{odd}} \rangle$ in the ground state of $H_{\text{Ryd}}$. This data is shown as the one with red circles in Fig. \ref{fig:transition}(a) and clearly displays a significantly sharper transition. Moreover, as shown in Fig. \ref{fig:transition}(b), one can anticipate a diverging behavior of the susceptibility $\chi_{\text{odd}}(\xi_s,\xi_d)$ with increasing lattice sizes.

This strongly suggests that $\{\xi_d=0, \xi_s^{-1} =0\}$ is a critical point of the Rydberg gas Hamiltonian (\ref{eqn:Ryd_Hamiltonian}) which is not captured by the ground state of the frustration-free approximation (\ref{eqn:R-K Hamiltonian}). To investigate the nature of this point we perform a scaling analysis by using the results shown in Fig. \ref{fig:transition}(b). Expressing the susceptibility as $\chi_{\text{odd}} \sim |\xi_d|^{-\gamma}$ with critical exponent $\gamma$, in the vicinity of $\xi_d=0$, we extract the critical exponent by fitting $\log{|\chi_{\text{odd}}|}$ as function of $\log{|\xi_d|}$ linearly within a linear scaling region \cite{go:92} and subsequently determining the gradient. For $\xi_s = 10$ and $L=18$, within an appropriately chosen scaling region, the critical exponent is found to be $\gamma \approx 1.76 \pm 0.05$. This result suggests that the second order transition belongs to the 2D-Ising universality class \cite{go:92}.

\begin{figure}
\includegraphics[width=0.85\columnwidth]{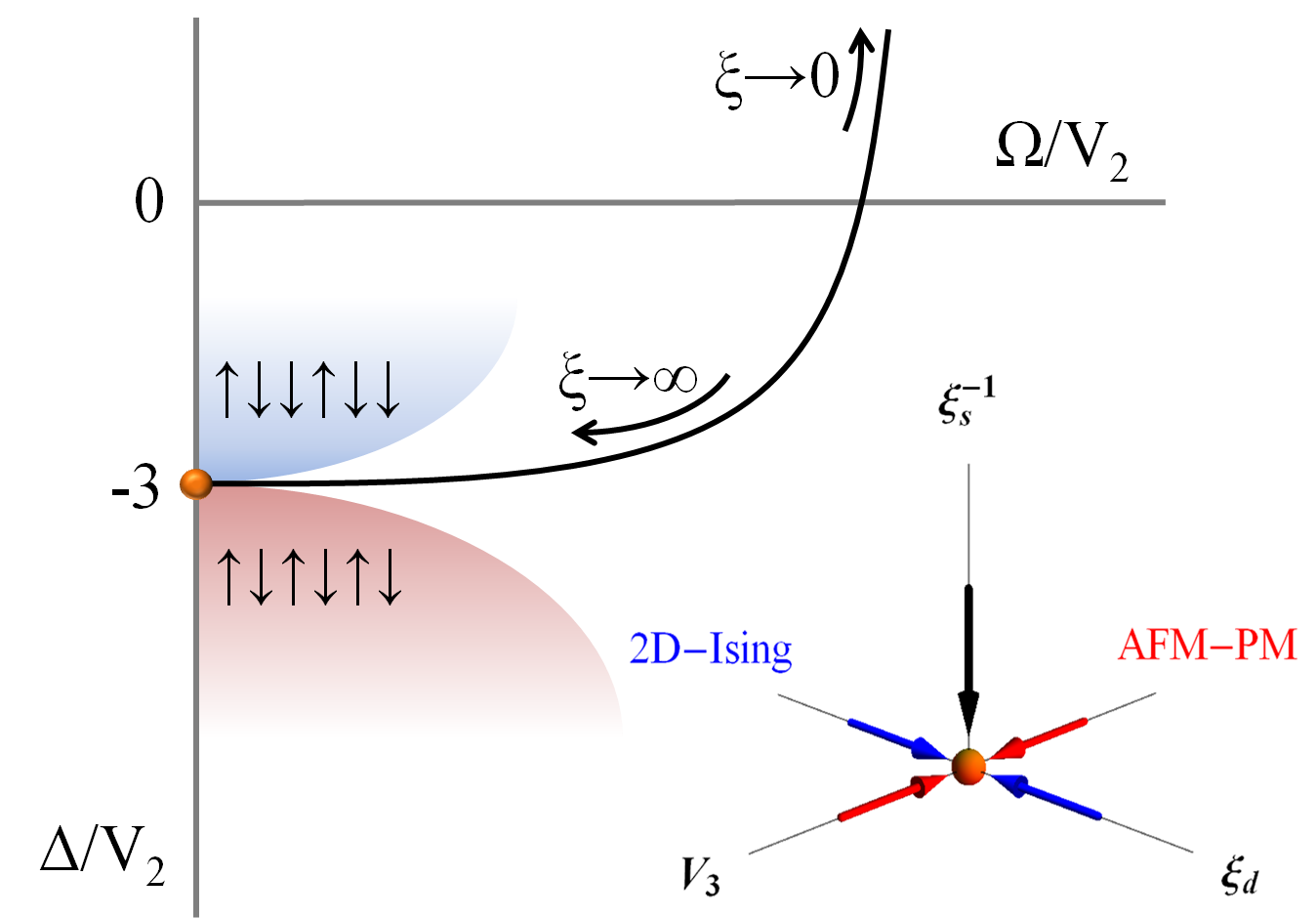}
\caption{Stylized phase diagram of the homogeneous Rydberg lattice gas Hamiltonian (\ref{eqn:Ryd_Hamiltonian}) with $\Omega_k=\Omega$ and $\Delta_k=\Delta$. The curve parameterized by Eqs. (\ref{eqn:RK-manifold}) approaches in the limit $\xi\to\infty$ the critical point $\{\Omega=0, \Delta=-3\, V_2\}$ which is located between a classical crystalline phase with Rydberg density $1/2$ and one with density $1/3$. Inset: In our analysis we approach the critical point $\{ \xi_d=0,\xi_s^{-1}=0, V_3=0\}$ from different directions: One experiences a 2D-Ising type transition by approaching the critical point from finite $\xi_d$, while one experiences a AFM-PM transition by approaching from finite $V_3$.}
\label{fig:phase_diagram}
\end{figure}

At the critical point we have $\xi_d=0$ and therefore $\xi_{\text{odd}}=\xi_{\text{even}}$ which means that the fugacity parameter is site independent, $\xi_k=\xi$. To understand the limits of the validity of the frustration-free Hamiltonain (\ref{eqn:R-K Hamiltonian}) and to gain further insights into the critical point it is therefore instructive to have a closer look at the phase diagram of the Hamiltonain (\ref{eqn:Ryd_Hamiltonian}) in  the homogeneous case. A stylized version of this diagram is depicted in Fig. \ref{fig:phase_diagram}. Here one sees that the curve, which is parameterized by Eqs. (\ref{eqn:RK-manifold}) hits the point $\{\Omega=0, \Delta_c=-3 V_2\}$, i.e. $\xi\rightarrow\infty$, where the phase boundaries of a classical phase with Rydberg density $1/3$ and one with density $1/2$ coalesce. An inspection of the frustration-free Hamiltonian (\ref{eqn:R-K Hamiltonian}) shows that in the limit $\xi\rightarrow\infty$ this approximate Hamiltonian does not energetically discriminate between configurations of the form $\left|\uparrow\downarrow\uparrow\downarrow\uparrow\downarrow...\right>$ and $\left|\uparrow\downarrow\downarrow\uparrow\downarrow\downarrow...\right>$ due to the lack of the next-next-nearest neighbor interaction $V_3$. This explains why its ground state (\ref{eqn:Ground state}) fails to describe properties, such as the sublattice density of the Rydberg lattice gas, in the vicinity of the critical point.

Having in mind that the presence of $V_3$ can affect the critical behavior, let us finally investigate in more detail the role of long-range interactions. To this end we amend the frustration-free Hamiltonian (\ref{eqn:R-K Hamiltonian}) by adding next-next-nearest neighbor interactions: $H_3=H_\mathrm{RK}+V_3\sum_k n_k n_{k+3}$. While in the previous inhomogeneous case, the critical point was approached from finite $\xi_d$ and growing $\xi=\xi_s/2$, we approach it now on the $V_3$ axis. This is depicted as the inset of Fig. \ref{fig:phase_diagram}.
We continue by deriving an effective Hamiltonian emerging from $H_3$ in the limit $\xi\rightarrow\infty$ following the procedure outlined in Refs. \cite{fese+:04,sepu+:11}: We introduce as new degrees of freedom the spin blocks $\mathbb{1} \equiv \uparrow\downarrow $ and $\mathbb{0} \equiv  \uparrow\downarrow\downarrow$. States composed out of a concatenation of these spin blocks are strictly degenerate under the frustration-free Hamiltonian (\ref{eqn:R-K Hamiltonian}) when $\xi\rightarrow\infty$. This degeneracy is broken by the next-next-nearest neighbor interaction $V_3$, and, at finite but large $\xi$, due to the presence of virtual transitions of the type $\uparrow\downarrow\downarrow\leftrightarrow\downarrow\uparrow\downarrow$.
In order to derive an effective Hamiltonian in this parameter regime one identifies the above-mentioned spin block $\mathbb{1}$ as fictitious hard core particle with creation operator $b^\dagger_k$, and the block $\mathbb{0}$ as corresponding fictitious hole. Following Refs. \cite{shre:80,sepu+:11} and introducing the spin operators $S_x^k=(b_k+b^\dagger_k)/2$, $S_y^k=i(b_k-b^\dagger_k)/2$ and $S_z^k=b^\dagger_kb_k-1/2$ one finds that the effective theory for large $\xi$ is given by the Heisenberg spin-1/2 XXZ-model in a magnetic field,
\begin{eqnarray*}
H_\mathrm{xxz}
=\sum_{j}\left[-\frac{1}{\xi}(S_x^j S_x^{j+1}+S_y^j S_y^{j+1})+\frac{\delta}{\xi} S_z^j S_z^{j+1}+\frac{\mu}{\xi} S_z^j\right].
\end{eqnarray*}
with $\delta=-1/3$ and $\mu=2/3+\xi\,V_3$. With respect to the pseudo-spin degree of freedom, i.e. $\left|\uparrow\right>_k$, and $\left|\downarrow\right>_k$, this model exhibits a critical transition between an anti-ferromagnetic (AFM) and a paramagnetic (PM) phase at $\mu=\delta+1$ \cite{yaya:66}, i.e. when $\xi\,V_3=0$. Hamiltonian $H_3$ has thus a critical point at $\{V_3=0, \xi^{-1}=0\}$. This corroborates the previous analysis of the numerical data displayed in Fig. \ref{fig:transition} which suggested a second order transition of the Rydberg lattice gas Hamiltonian (\ref{eqn:Ryd_Hamiltonian}) to occur at $\{\xi_d=0,\xi^{-1}_s=0\}$. In fact due to the finite nearest-neighbor and next-next-nearest neighbor interaction which is inevitably present in practice the Rydberg gas will never be strictly at the critical point. However, due to the smallness of $V_3$, stemming from the $r^{-6}$-scaling of the van der Waals interaction, observable exhibits a scaling behavior (as for example shown in Fig. \ref{fig:transition}) which will be interesting to be explored experimentally pursuing the route in Refs. \cite{wehe+:08,lowe+:09}.

In conclusion, we have presented an analysis of the static properties of a dense inhomogeneous Rydberg lattice gas based on a frustration-free Hamiltonian. This approach allowed us to analyze in detail the effect of an impurity on the correlation properties of the system. More importantly, the introduction of a symmetry breaking field allowed us to identify a critical point and to study the influence of long-range interactions on the critical properties of the system. In the future it will be interesting to extend this study to other inhomogeneous situations: For example, the generality of Hamiltonian (\ref{eqn:Ryd_Hamiltonian}) permits the exploration of disordered systems. Moreover, the fact that near the critical point the physics is effectively described by the XXZ-model might enable the experimental implementation and study of impurities immersed in Luttinger liquids along the lines of Ref. \cite{shmi+:97}.

\begin{acknowledgments}
\emph{Acknowledgements --- }
We gratefully acknowledge discussions with Hosho Katsura and Juan P. Garrahan and acknowledge funding from EPSRC and the ERA-NET CHIST-ERA (R-ION consortium). C.A. acknowledges support through a Feodor-Lynen Fellowship of the Alexander von Humboldt Foundation.
\end{acknowledgments}

\bibliographystyle{apsrev4-1}
\bibliography{isc}
\end{document}